\documentclass[usenatbib]{mn2e}

\usepackage[english]{babel}
\usepackage[latin1]{inputenc}
\usepackage[T1]{fontenc}

\usepackage{amsmath}
\usepackage{bbold}
\usepackage{graphicx}
\usepackage{amsfonts}
\usepackage{amssymb}
\usepackage{hyperref}
\usepackage{latexsym}

\newcommand{\be}{\begin{equation}}
\newcommand{\ee}{\end{equation}}
\newcommand{\bea}{\begin{eqnarray}}
\newcommand{\eea}{\end{eqnarray}}

\begin{document}


\title[Why multi-tracer surveys beat cosmic variance]{Why multi-tracer surveys beat  cosmic variance}

\author[L.~R.~Abramo \& K.~E.~Leonard]{L.~Raul Abramo 
$^{1}$\thanks{abramo@if.usp.br} and Katie E. Leonard $^2$\thanks{katie@phys.ufl.edu}
\\
$^1$ Departamento de F\'{\i}sica Matem\'atica,
Instituto de F\'{\i}sica, Universidade de S\~{a}o Paulo, 
CP 66318, CEP 05314-970 S\~{a}o Paulo, Brazil
\\
$^2$ Department of Physics, University of Florida, Gainesville, FL 32611, United States}

\maketitle

\begin{abstract}
Galaxy surveys that map multiple species of tracers of large-scale structure 
can improve the constraints on some cosmological parameters
far beyond the limits imposed by a simplistic interpretation of cosmic variance.
This enhancement derives from comparing the relative clustering between
different tracers of large-scale structure.
We present a simple but fully generic expression for the Fisher 
information matrix of surveys with any (discrete) number of tracers, and
show that the enhancement of the constraints on bias-sensitive parameters
are a straightforward consequence of this multi-tracer Fisher matrix. 
In fact, the relative clustering amplitudes between tracers are eigenvectors
of this multi-tracer Fisher matrix.
The diagonalized multi-tracer Fisher matrix clearly shows that while the 
effective volume is bounded by the physical volume of the survey, 
the relational information between species is unbounded.
As an application, we study the expected enhancements in the constraints 
of realistic surveys that aim at mapping several different types of tracers of 
large-scale structure. 
The gain obtained by combining multiple tracers is highest at low redshifts, and 
in one particular scenario we analyzed, the enhancement 
can be as large as a factor of $\gtrsim 3$ for the accuracy in the determination of the
redshift distortion parameter, and a factor $\gtrsim 5$ for the local non-Gaussianity 
parameter $f_{NL}$. Radial and angular distance determinations from the
baryonic features in the power spectrum may also benefit from the multi-tracer approach.
\end{abstract}



\section{Introduction}

Recent advances in Cosmology, in particular those related to the present epoch 
of accelerated expansion, have been overwhelmingly driven by 
astrophysical surveys \citep{York:2000gk,cole_2df_2005,Abbott:2005bi,scoville_cosmic_2007,adelman-mccarthy_sdss_2008,adelman-mccarthy_sixth_2008,PAN-STARRS,BOSS,2011MNRAS.415.2876B}.
The legacies of these surveys are a set of increasingly tight constraints on 
cosmological parameters and powerful tests of the robustness of the 
standard cosmological model, but also an alarming confirmation that 
there are gaping holes in that model.

The landscape of available instruments and observations is rapidly evolving. Besides 
general-purpose efforts that push the limits of what was achieved by the Sloan
Digital Sky Survey \citep{York:2000gk} both in wavelength and in redshift \citep{BigBOSS,SUMIRE,PFS}, 
now cosmologists can also count on surveys with a cadence aimed at 
discovering variable objects \citep{PTF,LSST:2009pq}, and 
mapping large volumes of the 
universe with high completeness. The latter is achieved by imaging large 
areas of the sky with narrow-band filters \citep{Benitez:2008fs} or
resorting to low-resolution (spectral as well as spatial) integral-field 
spectroscopy \citep{HETDEX}.

Galaxy redshift surveys are now expected to perform high-accuracy 
measurements of the equation of state of dark energy, test for models 
of modified gravity through lensing or redshift-space distortions \citep{LinderGrowth}, impose constraints on 
the amplitude of non-Gaussian features in the power spectrum \citep{NG2000,BartoloRev}, 
 measure neutrino masses, etc. -- see, e.g. \citep{Benitez:2008fs,SUMIRE,PFS}. 
Since many surveys are already reaching large fractions of the sky with 
high completeness, the variance due to the survey's finite volume (i.e. cosmic 
variance) is perhaps the most formidable obstacle to further progress.

Interestingly, it has been pointed out by
\citet{SeljakNG,McDonald:2008sh} 
that cosmic variance can be somehow circumvented for some cosmological 
parameters -- see also \citet{GM2010,CaiCV,Hamaus2011,Hamaus2012} . 
This is true, in particular, for the 
redshift-space distortion (RSD) parameter, $\beta(z)=f(z)/b(z)$, and for the 
amplitude of local non-Gaussianities, $f_{NL}$. The reason behind this ``miracle''
is that bias-sensitive 
parameters such as $\beta$ and $f_{NL}$ are not subject to the same random
processes that lead to different realizations of the density field for some 
matter power spectrum $P(k)$. Therefore, given a fixed density field, by comparing the clusterings between different types of tracers of large-scale structure (i.e. objects
which correspond to halos of different masses) we should be 
able to measure these parameters with a precision that is not limited 
by cosmic variance. We should point out that the source of this extra information
is not some sub-Poissonian shot noise that can be achieved by, for example,
mass-weighting instead of bias-weighting \citep{Seljaketal09,2011MNRAS.412..995C},
but is rather a direct consequence of the different relative amplitudes of 
the clustering among the tracers.

Using these ideas, \citet{Hamaus2011,Hamaus2012},
constructed a covariance matrix for the ratios of the clusterings of any two types of 
tracers, and used N-body simulations to obtain enhanced constraints for 
$\beta$ and $f_{NL}$. 
Conversely, \citet{Slosar09} constructed the Fisher information for $f_{NL}$
directly from the covariance matrix for counts in cells under simplifying 
assumptions similar to ours.
\citet{GM2010}, on the other hand, studied how non-linear bias and bias 
stochasticity can degrade the two-tracer constraints, and found that
the actual improvements in the constraints may be up to $\sim$ 50\% 
smaller, compared with the ideal case of linear and deterministic bias. 
\citet{CaiCV} pointed out that bias modulations between dark matter halos
imply correlations between the biases of the tracers, which limits the 
potential gains of the multi-tracer approach. \citet{CaiWL} then analyzed 
how weak gravitational lensing tomography surveys can be combined with
redshift surveys in order to mitigate those correlations, and to break the 
degeneracies between the biases and the other cosmological parameters of 
interest.


In this paper we show that the enhanced constraints of bias-sensitive
parameters in multi-tracer surveys are a direct consequence of the 
Fisher matrix for multiple tracers of large-scale structure. The analysis is 
an extension of the single-tracer Fisher matrix of \citet{1994ApJ...426...23F} 
(henceforth FKP) to multiple tracers of large-scale structure. What emerges 
from the multi-tracer Fisher matrix is, besides the usual  
cosmic variance-limited ``effective volume'' 
\citep{Tegmark_Surveys_1997,1998ApJ...499..555T}, 
simple expressions which quantify the amount of information that lies in 
the ``relational" degrees of freedom, which are completely independent 
of the effective volume. 

We show that there is a simple choice of variables which diagonalizes the
multi-tracer Fisher matrix. The total effective volume of the survey then appears 
naturally as one of the terms of this diagonal matrix, while the remaining terms 
correspond to the information in the relative clusterings between the different 
species of tracers. The key aspect is that the Fisher matrix elements
corresponding to the relational degrees of freedom are unbounded, and
can in fact carry much more information about some parameters than the
effective volume -- which is, of course, bounded by the physical volume 
of the survey.

This paper is organized as follows: in Sec. \ref{S:1} we review the statistics
of galaxy surveys, and present the generalization of the FKP method to multiple
tracers; in Sec. \ref{S:2} we show how a certain choice of relative clusterings
between the tracers leads to the diagonalization of the multi-tracer Fisher matrix;
 in Sec. \ref{S:3} we show an application to a hypothetical survey of three
types of tracers, and how the relational information improves the constraints; our conclusions comprise Sec. \ref{S:4}.

\section{The statistics of galaxy surveys}
\label{S:1}

The main observables galaxy surveys try to measure
are the redshift-space matter power spectrum and its 
sub-products, such as the
baryon acoustic oscillations \citep{eisenstein_cosmic_1998,blake_probing_2003,
seo_probing_2003} and redshift-space distortions 
\citep{HamiltonRev05a,HamiltonRev05b}.
In terms of the matter density contrast $\delta(\vec{x})$, the position-space power 
spectrum is given by the expectation value 
$\langle \delta (\vec{k},z) \delta^*(\vec{k}',z) \rangle = (2\pi)^3 P(k,z) 
\delta_D (\vec{k}-\vec{k}')$, where $\delta_D$ is the Dirac delta function.

The problem with this program is that we cannot directly measure the 
fluctuations of the matter field in position space. Instead, what we observe are counts 
of tracers of the large-scale structure in redshift space. In the simplest 
approximation, bias and redshift distortions can be regarded as operators acting 
on the density contrast, such that the fluctuation field of some tracer $\alpha$ 
(where $\alpha$ stands for galaxies of a certain type) 
in redshift space is related to the underlying mass fluctuation field in position space 
by some relation
$ \delta_\alpha (\vec{k},z) \sim [b_\alpha + f \, \mu_k^2] \delta(\vec{k},z) \; $.
Here $\delta_\alpha=(n_\alpha-\bar{n}_\alpha)/\bar{n}_\alpha$ is the fluctuation in the 
number density of the tracer species $\alpha$ over the average ($\bar{n}_\alpha$)
in redshift space, $b_\alpha$ is the bias of that tracer, 
$f=d\log G/d\log a$ is the RSD parameter ($G$ is the matter
growth function, and $a$ the scale factor), and $\mu_k=\hat{k}\cdot\hat{r}$ 
is the cosine of the angle of the Fourier mode with the line of sight.

As a first approximation, we can assume the bias to be linear, 
deterministic, and scale-independent \citep{BBKS}. However, it is known that structure formation leads to scale-dependent, non-linear and stochastic
bias \citep{Benson:1999mva,BiasLahav,BiasWeinberg,BiasRavi}. Moreover, 
even the primordial spectrum of cosmological fluctuations may effectively introduce
a scale-dependent bias through non-Gaussian features
\citep{BartoloRev,Sefusatti2007,NGDalal}. Hence, we can regard the bias
as being not only a function of redshift, but a function of scale, having
at least some stochastic component. The RSD parameter and the angular
dependence can also inherit scale-dependent non-linear corrections
\citep{Raccanelli2012}. Since the power spectrum, bias, RSDs, and even the 
selection functions are estimated 
from the same data, it is important to understand 
how accurately one can measure these quantities and how their covariances carry 
over to the uncertainties in the parameters of interest. 
In the absence of data, the best tool for studying these issues is the Fisher 
information matrix.

\subsection{The Fisher matrix}

The main sources of uncertainty in galaxy surveys are cosmic variance,
which is due to the finite volume of the survey, and shot noise, which arises
from the statistics of the counts of tracers. 
In the case of surveys with a single species of tracer,
the optimal weighting function for galaxy counts, which minimizes the joint contributions
of cosmic variance and shot noise to the variance of the power spectrum estimates, 
was first obtained by 
\citet{1994ApJ...426...23F}. The corresponding Fisher information matrix
was derived by \citet{Tegmark_Surveys_1997,1998ApJ...499..555T}, who also 
showed that this Fisher matrix follows from the covariance of counts of galaxies
\citep{1998ApJ...499..555T}.
The FKP Fisher matrix for the survey of the tracer species
$\alpha$ can be written as:
\be
\label{Eq:FKP_Fisher}
F_{\alpha; \, ij} = \int \frac{d^3 k \, d^3 x}{(2\pi)^3}
\, \frac{d \log {\cal{P}}_\alpha}{d \theta^i} 
\, \frac12 \left( \frac{{\cal{P}}_\alpha}{1+{\cal{P}}_\alpha} \right)^2
\, \frac{d \log {\cal{P}}_\alpha}{d \theta^j} \; ,
\ee
where we define: 
\be
\label{Eq:X}
{\cal{P}}_\alpha(\vec{k};\vec{x}) :=  \bar{n}_\alpha(\vec{x}) P_\alpha (\vec{k}; \vec{x}) \; .
\ee
The dimensionless quantity ${\cal{P}}_\alpha$, which we 
will call the {\em effective power},
expresses the power spectrum of the tracer $\alpha$ in units of its
shot noise (which we take to follow the standard Poisson distribution, 
$1/\bar{n}_\alpha$). Under the usual assumptions about bias 
and the nature of RSDs, the effective power can be expressed as:
\be
\label{Eq:X_alpha}
 {\cal{P}}_\alpha (\vec{k};\vec{x}) \rightarrow 
\bar{n}_\alpha(\vec{x}) \left[ b_\alpha + f(z) \mu^2_k \right]^2 P(k,z) \; .
\ee

Equation (\ref{Eq:FKP_Fisher}) shows that we can regard the quantity:
\be
\label{Eq:F_alpha}
F_\alpha (\vec{k};\vec{x}) = \frac12 \left( \frac{{\cal{P}}_\alpha}{1+{\cal{P}}_\alpha} \right)^2 \; ,
\ee
as the Fisher information density in phase space. In fact, the 
$F_\alpha(\vec{k};\vec{x})$ defined above is the Fisher information matrix for the 
parameter $\log {\cal{P}}_\alpha(\vec{k};\vec{x})$, and Eq. \eqref{Eq:F_alpha}
shows that the Fisher information density per unit of phase space volume 
is limited by $F_\alpha < \frac12$.
Therefore, for a survey of a single species of tracer there is an absolute limit on the 
amount of information that can be extracted 
from a finite volume and a finite range of scales. That upper limit is precisely 
$\frac12$ for each volume element of phase space $d^3 k \, d^3 x / (2\pi)^3 $.
Cosmic variance is just the position-space manifestation of this upper limit.

The Fisher matrix of Eq. (\ref{Eq:FKP_Fisher}) can be used to estimate
all sorts of parameters, including the underlying matter power spectrum.
For estimations of the power spectrum, we typically assume that the
spectrum of the tracer is proportional to that of matter, and take
the parameters $\theta^i \rightarrow p^i = \log P(k_i)$, where the $k_i$
are bins in Fourier space. In that case,
it is trivial to see that Eq. (\ref{Eq:FKP_Fisher}) reduces to:
\be
\label{Eq:FKP_P}
F_{\alpha; \, i j}  \rightarrow \delta_{i j} \times \, \frac12 \, \frac{4\pi k_i^2 \Delta k_i}{(2\pi)^3}
\int d^3 x
\, \left( \frac{{\cal{P}}_\alpha}{1+{\cal{P}}_\alpha} \right)^2
\; ,
\ee
where the integral over position space defines the usual {\em effective volume} 
\citep{Tegmark_Surveys_1997,1998ApJ...499..555T}. Hence, 
the uncertainty in the amplitude of the power spectrum at the scale $k_i$
is given by $ \sigma^2 (p^i) = \sigma^2_{P_i}/P_i^2 = \frac 12 V_{k_i} \, V_{eff} (k_i)$.

\subsection{Multi-species Fisher matrix}

When several types of tracers are observed over the same 
volume cosmic variance should remain unaffected, but the nature of shot noise 
means that counts of one type of galaxy affect the counts of 
the other types in a Fisher information matrix.
The optimal (minimal-variance) estimator
in the case of multiple tracers was first obtained by \citet{Percival:2003pi},
who considered a continuous distribution of galaxies with 
luminosity-dependent bias. Those results were later used to write a Fisher matrix 
for surveys of multiple species \citep{White:2008jy,McDonald:2008sh}.

In \citet{AbramoFisher}, it was shown that the multi-tracer Fisher matrix also
follows from the covariance of counts of galaxies, under the usual assumptions
and approximations -- e.g., that shot noise does not apply to the
cross-correlations, see \citet{Smith09}. The results of that paper 
are summarized below.

Let's define the {\em total effective power} as the sum of the effective powers
of all $N_t$ species of tracers in a survey:
\be
\label{Eq:def_X}
{\cal{P}} (\vec{k};\vec{x}) := \sum_{\alpha=1}^{N_t} {\cal{P}}_\alpha \; .
\ee

The Fisher matrix for the parameters $\log {\cal{P}}_\alpha (\vec{k};\vec{x})$
and $\log {\cal{P}}_\beta (\vec{k};\vec{x})$ is then given by \citep{AbramoFisher}
\be
\label{Eq:F_X}
F_{\alpha \beta} (\vec{k};\vec{x}) = 
\frac14 
\left[ \delta_{\alpha \beta}  \, \frac{{\cal{P}}_\alpha  \, {\cal{P}}}{1+{\cal{P}}} + 
\frac{{\cal{P}}_\alpha  \, {\cal{P}}_\beta  \, (1-{\cal{P}})}{(1+{\cal{P}})^2} \right] \; .
\ee

The Fisher matrix defined in Eq. \eqref{Eq:F_X} can be obtained from that of 
\citet{White:2008jy} by a simple projection. \citet{White:2008jy} do not
assume any relationship between the cross-power spectra of two species,
$P_{\alpha\beta}$, and their auto-power spectra. We, on
the other hand, work under the assumption that bias stochasticity
vanishes for the relevant scales, so that the cross-power 
spectra are implicitly given in terms of the auto-power spectra through 
relations such as $P_{\alpha\beta}^2 = P_{\alpha\alpha} P_{\beta\beta}$.
-- see, however, \citet{Swansonetal08} and \citet{Bonoli} for the limitations
of this approach. 
By applying this restriction to the Fisher matrix for {\em pairs} 
of tracers in \citet{White:2008jy}, we obtain the Fisher matrix of 
Eq. \eqref{Eq:F_X}.

In terms of a more usual set of parameters, $\theta^i$, we have:
\bea
\label{Eq:F_X_ij}
F_{ij} &=& \sum_{\alpha \beta} \int \frac{ d^3 k \, d^3 x}{(2\pi)^3}
\frac{d \log {\cal{P}}_\alpha}{d \theta^i}
F_{\alpha \beta} 
\frac{d \log {\cal{P}}_\beta}{d \theta^j} 
\\ \nonumber
& := &  \sum_{\alpha \beta} F_{\alpha \beta ; \, i j} \; .
\eea
Notice that the Fisher information density is symmetric on the tracer
species indices, $F_{\alpha \beta}(\vec{k};\vec{x}) = F_{\beta \alpha}(\vec{k};\vec{x}) $, 
but, from the definition above, 
$F_{\alpha \beta; ij} = F_{\beta \alpha; j i} \neq F_{\beta \alpha; i j} $.

Eq. \eqref{Eq:F_X_ij}  shows, as was already the case for Eq. \eqref{Eq:F_alpha}, 
that we ought to call $F_{\alpha\beta}$ 
the {\em multi-tracer Fisher information density} per unit of 
phase space volume. In the case of a single species of tracer
Eq. \eqref{Eq:F_alpha} tells us that the information density in phase space has
an upper bound of $\frac12$; for multiple tracers Eqs. \eqref{Eq:F_X} and 
\eqref{Eq:F_X_ij} reveal that the information density is unbounded. 

In the case of two tracers, the Fisher matrix for the parameters
$\log {\cal{P}}_1(\vec{k};\vec{x})$ and $\log {\cal{P}}_2(\vec{k};\vec{x})$  is given by
\be
\label{Eq:F12}
F_{\alpha \beta} =
\frac14 \left(
\begin{array}{lll}
\frac{{\cal{P}}_1 \, {\cal{P}}}{1+{\cal{P}}} +
\frac{{\cal{P}}_1^2 (1-{\cal{P}})}{(1+{\cal{P}})^2}
 & 
 {} 
 &
 \frac{{\cal{P}}_1 {\cal{P}}_2 (1-{\cal{P}})}{(1+{\cal{P}})^2}
 \\
 {} & {} \\
 \frac{{\cal{P}}_1 {\cal{P}}_2 (1-{\cal{P}})}{(1+{\cal{P}})^2}
 & 
 {} 
 & 
 \frac{{\cal{P}}_2 \, {\cal{P}}}{1+{\cal{P}}} +
\frac{{\cal{P}}_2^2 (1-{\cal{P}})}{(1+{\cal{P}})^2}
\end{array}
\right) \; .
\ee
Either from Eq. \eqref{Eq:F_X} or from Eq. \eqref{Eq:F12} it can be seen that
the individual elements of the multi-tracer Fisher matrix are unbounded, in
contrast to the single-species Fisher matrix of Eq. \eqref{Eq:F_alpha}. Nevertheless, as we will see next, 
cosmic variance is still manifested in the multi-tracer Fisher matrix.

We can readily obtain the FKP Fisher matrix for the power spectrum from
Eq. \eqref{Eq:F_X_ij} by noting that, if the $\theta^i$ are limited to parameters
of the power spectrum, $p^i$, which are shared equally by all tracers, then 
$d \log {\cal{P}}_\alpha/d\theta^i \rightarrow d \log P/dp^i$.
In that case, the sum over tracers in Eq. \eqref{Eq:F_X_ij} can be brought
inside the integral, resulting in:
\be
\label{Eq:def_Ft}
F_T :=
\sum_{\alpha \beta} F_{\alpha \beta} = \frac12 \left( \frac{{\cal{P}}}{1+{\cal{P}}} \right)^2 
 < \frac12 \; .
\ee
This then leads to the familiar expression for the Fisher matrix, 
shown in Eq. \eqref{Eq:FKP_P}, 
provided we make the natural identification 
${\cal{P}}_\alpha \rightarrow {\cal{P}} = \sum_\alpha {\cal{P}}_\alpha$. 
This result means that, when measuring $P(k,z)$,
we are always limited by cosmic variance 
(through the effective volume), and shot noise is 
determined by the sum of the effective powers. Under the
usual assumptions, this total effective power reduces to 
${\cal{P}} \rightarrow \left[ \sum_\alpha \bar{n}_\alpha (b_\alpha + f \, \mu_k^2)^2 \right] P(k,z)$.

However, even though the Fisher information density
for the power spectrum has an upper limit,
$F_T  < \frac12$, the individual components
of the Fisher matrix density, $F_{\alpha\beta}$, are not bounded. 
In fact, this upper limit is only relevant in the case of
parameters for which the derivatives $d \log {\cal{P}}_\alpha/d\theta^i$ are
{\em independent} of the tracer species $\alpha$. This is the case, e.g., for
the power spectrum and for the matter growth function $G(z)$, 
but is {\em not} the case for the biases of the tracers, for the RSD parameter, 
or for any other bias-sensitive parameter such as $f_{NL}$.

In particular, this means that the Fisher information density for some 
parameters can be much larger than the na\"{\i}ve $\frac12$ bound.
Indeed, inspection of Eq. \eqref{Eq:F_X} shows that the diagonal components 
of $F_{\alpha \beta}$ can be arbitrarily large when some of the effective
powers ${\cal{P}}_\alpha \gg 1$. The cross-terms ($\alpha \neq \beta$),
on the other hand, can be very large and negative, but this is precisely 
what is necessary in order to preserve the constraint of Eq. \eqref{Eq:def_Ft}.

\section{Diagonalized multi-species Fisher matrix}
\label{S:2}

Even if some individual components of the multi-tracer Fisher matrix
$F_{\alpha \beta}$ are arbitrarily
large, this does not necessarily mean that the Fisher matrix for the parameters,
shown in Eq. \eqref{Eq:F_X_ij}, will inherit these enhancements after summing
over all the tracers and accounting for the cross-correlations. 
We will now show that this is in fact the case, by diagonalizing
the multi-tracer Fisher matrix.

The parameters of the multi-tracer Fisher matrix in Eq. \eqref{Eq:F_X} are
the variables $\log {\cal{P}}_\alpha (\vec{k};\vec{x})$ ($\alpha=1,\ldots,N_t$). 
Let's define, in a manner similar to Eq. \eqref{Eq:def_X}, the partial sums:
\be
\label{Eq:def_Xb}
{\cal{S}}_a = \sum_{\alpha=a}^{N_t} {\cal{P}}_\alpha \; ,
\ee
where $a=1,\ldots,N_t$, and by this definition ${\cal{S}}_1={\cal{P}}$
and ${\cal{S}}_{N_t}={\cal{P}}_{N_t}$.
The variables ${\cal{S}}_a$ can be regarded as the {\em aggregate} effective powers.
The order in which the effective powers ${\cal{P}}_\alpha$ are organized
in these sums is irrelevant.

The set of variables which diagonalizes the multi-tracer Fisher matrix are 
defined as follows:
\bea
\label{Eq:def_Y1}
{\cal{Y}}_1 & = & {\cal{S}}_1 = {\cal{P}} \; ,
\\
\nonumber
{\cal{Y}}_2 & = & \frac{{\cal{P}}_1}{{\cal{S}}_2} \; ,
\\
\nonumber
&\vdots&
\\
\label{Eq:def_Ya}
{\cal{Y}}_{a} & = & \frac{{\cal{P}}_{a-1}} {{\cal{S}}_{a}}
\quad (a \neq 1) \; ,
\\
\nonumber
&\vdots&
\\
\label{Eq:def_YN}
{\cal{Y}}_{N_t} & = & \frac{{\cal{P}}_{N_t-1}} {{\cal{S}}_{N_t}}
= \frac{{\cal{P}}_{N_t-1}}{{\cal{P}}_{N_t}} \; .
\eea
Hence, the new variable ${\cal{Y}}_1$ is simply the
total effective power of the survey, while the other variables ${\cal{Y}}_a$ ($a\neq 1$) 
are {\em relative} effective powers. These relative powers are
ratios of the individual effective powers to the aggregate effective 
powers ${\cal{S}}_a$.
While any ordering of the effective powers can be used for defining the variables ${\cal{S}}_a$ and 
${\cal{Y}}_a$, it may be more intuitive to organize the tracers in order of
the volume that they cover in the survey, so that the tracer with the least volume would 
correspond to ${\cal{P}}_1$ and the tracer covering the most volume would 
correspond to ${\cal{P}}_{N_t}$. However, one may also organize the effective powers
as a function of bias (or halo mass). In either case, the Fisher matrix always remains 
well-behaved, even if some of the ${\cal{P}}_\alpha$ happen to vanish.

Changing variables from $\log {\cal{P}}_\alpha (\vec{k};\vec{x})$
to $\log {\cal{Y}}_a (\vec{k};\vec{x})$ leads to a completely diagonal
Fisher matrix:
\bea
\nonumber
F_{ab} &=& \sum_{\alpha\beta} 
\frac{d \log {\cal{P}}_\alpha}{d \log {\cal{Y}}_a}
F_{\alpha\beta} 
\frac{d \log {\cal{P}}_\beta}{d \log {\cal{Y}}_b}
\\
\label{Eq:F_Y}
&=& \delta_{ab} \, {\cal{F}}_a \; .
\eea
The elements of this diagonal Fisher matrix are:
\bea
\label{Eq:f_1}
{\cal{F}}_1 &=& \frac12 \left( \frac{{\cal{P}}}{1+{\cal{P}}} \right)^2 = F_T
\\
\label{Eq:f_a}
{\cal{F}}_a &=& 
\frac14 \frac{{\cal{P}}}{1+{\cal{P}}} \frac{ {\cal{S}}_a \, {\cal{P}}_{a-1}}{{\cal{S}}_{a-1}}
 \quad (a\neq 1) \; .
\eea

Therefore, the variables $\log {\cal{Y}}_a$, with the definitions of 
Eqs. \eqref{Eq:def_Y1}-\eqref{Eq:def_Ya}, are the eigenvectors of the multi-tracer 
Fisher matrix, and its eigenvalues are the ${\cal{F}}_a$ of Eqs. 
\eqref{Eq:f_1}-\eqref{Eq:f_a}.

It follows that the variable ${\cal{Y}}_1={\cal{P}}$ (the total effective
power of the survey) is completely independent from the relational variables 
${\cal{Y}}_a$ ($a\neq1$), and, by Eq. \eqref{Eq:f_1}, its information density 
is strictly limited by the constraint $0 \leq {\cal{F}}_1<\frac12$. 
The variables ${\cal{Y}}_a$ ($a\neq1$) are also independent of each other, 
but from Eq. \eqref{Eq:f_a} the information densities associated with 
them can assume any (positive) value. Since the multi-tracer Fisher matrix is diagonal, the 
uncertainties in $\log {\cal{Y}}_a$ are given by the square roots of the
inverses of Eqs. \eqref{Eq:f_1}-\eqref{Eq:f_a}. 

Any linear combination of the variables $\log {\cal{Y}}_a$ constructed through the action 
of an $N_t$-dimensional orthogonal transformation would still lead to a diagonal 
Fisher matrix. It can be verified that permutations of the effective 
powers ${\cal{P}}_\alpha$ generate ${\rm O}(N_t-1)$ orthogonal transformations 
between the variables $\log {\cal{Y}}_a$ ($a\neq 1$) which are equivalent
to the corresponding redefinitions according to Eqs. \eqref{Eq:def_Y1}-\eqref{Eq:def_Ya}.

As a concrete example, take a survey of two species of tracers. 
In that case, we have ${\cal{Y}}_1={\cal{P}}_1+{\cal{P}}_2={\cal{P}}$,
and ${\cal{Y}}_2={\cal{P}}_1/{\cal{P}}_2$. The Fisher matrix element associated with 
$\log {\cal{Y}}_1$ is ${\cal{F}}_1=\frac12 \, {\cal{P}}^2/(1+{\cal{P}})^2$, and the Fisher 
matrix for  $\log {\cal{Y}}_2$ is ${\cal{F}}_2=\frac14 \, {\cal{P}}_1 {\cal{P}}_2/(1+{\cal{P}})$.
Exchanging ${\cal{P}}_1$ and ${\cal{P}}_2$ leaves $\log {\cal{Y}}_1$ invariant,
introduces an irrelevant change in the sign of $\log {\cal{Y}}_2 \rightarrow -\log {\cal{Y}}_2$,
and leaves both ${\cal{F}}_1$ and ${\cal{F}}_2$ invariant.

\begin{figure}
\includegraphics[width=80mm]{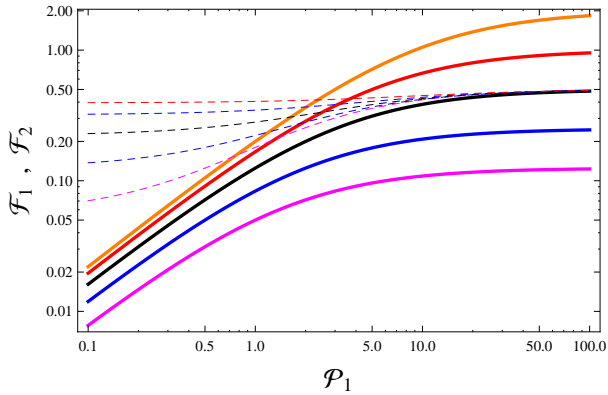}
\caption{Fisher matrix elements in the case of two species of tracers. 
The Fisher information density ${\cal{F}}_1$ of Eq. \eqref{Eq:f_1}, 
which is associated with the total effective spectrum
${\cal{Y}}_1={\cal{P}}={\cal{P}}_1+{\cal{P}}_2$, is shown by the dashed lines
for various values of ${\cal{P}}_2$ (0.5, 1, 2, 4, and 8, from the bottom up).
The Fisher information density 
${\cal{F}}_2=\frac14 {\cal{P}}_1 {\cal{P}}_2/(1+{\cal{P}}_1+{\cal{P}}_2)$, associated with
the relative power ${\cal{Y}}_2={\cal{P}}_1/{\cal{P}}_2$, is shown by the solid lines.
}
\label{Fig:f12}
\end{figure}

The behavior of the two independent components of the Fisher information density
for the two-tracer case are plotted in Fig. \ref{Fig:f12}. For small values of
the effective powers ${\cal{P}}_1$ and ${\cal{P}}_2$, it is ${\cal{F}}_1$ which
has the largest information density. However, for large values of the
effective power (that is, for large enough densities of the tracers), it
is ${\cal{F}}_2$ which carries the most information density.
In the limit ${\cal{P}}_1 \gg 1$ we see that
${\cal{F}}_1 \rightarrow \frac12 $, while ${\cal{F}}_2 \rightarrow \frac14 {\cal{P}}_2$;
hence, when both ${\cal{P}}_1 \gg 1$ and ${\cal{P}}_2 \gg 1$, we have
${\cal{F}}_1 \approx \frac12$, but ${\cal{F}}_2 \gg 1$. If 
${\cal{P}}_1$ or ${\cal{P}}_2$ vanishes in some region of space, this region 
will not contribute with any information about their ratio (${\cal{Y}}_2$), 
although it does contribute to the usual Fisher matrix density (the one
associated with ${\cal{Y}}_1={\cal{P}}$). When both tracers vanish in some
region of space, then the whole Fisher matrix also vanishes identically.

This diagonal form of the Fisher matrix has an additional advantage: it 
reduces the amount of computations needed for practical applications. 
Instead of the $N_t(N_t-1)/2$ sums and integrations in Eq. \eqref{Eq:F_X_ij}, 
we only need to compute $N_t$ terms:
\be
\label{Eq:F_Y_ij}
F_{i  j} = \sum_{a=1}^{N_t} \int \frac{d^3 k \, d^3 x}{(2\pi)^3} 
\, \frac{d \log {\cal{Y}}_a}{d \theta^i} \,
{\cal{F}}_{a} \,
\frac{d \log {\cal{Y}}_a}{d \theta^j} \; .
\ee
This means if it ever becomes possible (and desirable) to divide the tracers in a survey 
into $100$ different types, we only need to compute the 100 terms in 
Eq. \eqref{Eq:F_Y_ij}, instead of the $\sim 5.\times 10^3$ terms needed for the 
non-diagonal form of the Fisher matrix.

\vskip 0.3cm

\section{Applications to future surveys}
\label{S:3}

As an application of these results, we
study how the relative clusterings improve cosmological
constraints for a hypothetical redshift survey that can
detect three types of tracers of large-scale structure.
These tracers were chosen to reproduce, as much as possible,
the properties of luminous red galaxies (LRGs), emission-line
galaxies (ELGs) and quasars or AGNs (QSOs). The LRG-like tracers
are relatively rare, have a somewhat high bias, and are shallow 
($z\lesssim1.5$). The ELG-like tracers are more abundant, have a 
relatively low bias, and can be detected to higher redshifts compared
to LRGs ($z \lesssim 2$). The QSO-like tracers are very rare, 
have a very high bias, and can be detected to very high redshifts
($z \lesssim 4$) -- see, e.g., \citet{2011arXiv1108.2657A}.

\begin{figure}
\includegraphics[width=80mm]{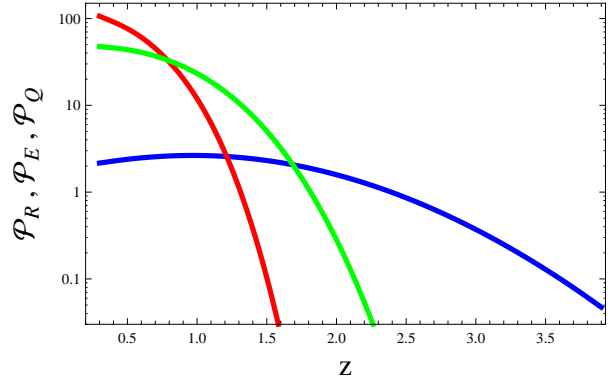}
\caption{Effective powers ${\cal{P}}_R$ (LRG-like, red in color version), 
${\cal{P}}_E$ (ELG-like, yellow), and ${\cal{P}}_Q$ (QSO-like, blue),
evaluated at $k=0.1 \, h \, {\rm Mpc}^{-1}$, and across the line-of-sight ($\mu_k=0$).
}
\label{Fig:Effpots}
\end{figure}

Fig. \ref{Fig:Effpots} shows the effective powers for each species of
tracer, computed at the typical scale of $k=0.1$ $h$ Mpc$^{-1}$,
for modes perpendicular to the line-of-sight ($\mu_k=0$). Since the effective
power is a measure of shot noise (a high value of ${\cal{P}}_\alpha$ indicates
very low shot noise), the effective powers chosen for Fig. \ref{Fig:Effpots}
cover several different scenarios that one may encounter in real surveys.

We have assumed that the survey covers $10^4$ deg$^2$, which, for the number
densities we have considered, imply total numbers of $2 \times 10^7$ for the
LRG-like tracers, $5 \times 10^7$ for the ELG-like tracers, and $3 \times 10^6$ 
for the QSO-like tracers. 
We also assumed that the redshifts are accurate to $\sigma_z = 0.001 (1+z)$,
and these uncertainties were factored into the Fisher matrix in the usual way, 
through a factor $\exp [ - k^2 \, \mu_k^2 \, \sigma_z^2 \, c^2 \, H^{-2} ]$ which 
multiplies the Fisher information density.
We have also cut-off the Fourier-space
integrations at $k=0.1$ $h$ Mpc$^{-1}$, in order to avoid 
contributions from scales where non-linear effects become essential.

We employ the same method that has been extensively used to make forecasts
using the power spectrum and baryonic acoustic oscillations (BAOs)
\citep{seo_probing_2003,2006ApJ...647....1W}.
We assume that the observed power spectrum for a given tracer
can be expressed in terms of a ``reference'' model as:
\be
\label{Eq:ObsPk}
P_{obs}(z;k^{ref},\mu_k^{ref}) = \left[ \frac{D_A^{ref}(z)}{D_A(z)} \right]^2
\frac{H(z)}{H^{ref}(z)} \, P_{l}(z;k^{ref},\mu_k^{ref}) \; ,
\ee
where:
\be
\label{Eq:Plin}
P_{l}(z;k,\mu) = G^2(z) \left[ b(z) + f (z) \, \mu^2_k \right]^2 \, P_{l}(0;k) \; .
\ee
In this last equation, $P_l(0,k)$ denotes the linear theory, position 
space power spectrum at $z=0$.
In Eq. (\ref{Eq:ObsPk}) the wavenumbers in the fiducial model 
are related to those in another, arbitrary cosmology, by:
\be
\label{Eq:ks}
k^{ref}_{||} = k_{||} \frac{H^{ref}(z)}{H(z)} \quad , \quad
k^{ref}_{\perp} = k_{\perp} \frac{D_A(z)}{D_A^{ref}(z)} \; .
\ee
Changes in the properties of $k$ and $\mu_k$ due to changes in the cosmological model 
follow from their definitions,
$k=\sqrt{k_{||}^2 + k_{\perp}^2}$, and $\mu^2_k = k_{||}^2/k^2$.
For simplicity, we will follow \citet{2006ApJ...647....1W} and 
absorb the normalization of the power spectrum, as well as the matter 
growth function, inside the prefactor of Eq. \eqref{Eq:Plin}. 
This leads to the following spectrum for
the species $\alpha$:
\be
\label{Eq:Plin2}
P_{\alpha}(z;k,\mu_k) = \left[ s_\alpha(z) + f_s (z) \, \mu^2_k \right]^2 
\, \frac{P_{l}(0;k)}{\sigma_8^2} \; ,
\ee
where the {\em effective bias} for the tracer $\alpha$ is defined as:
\be
\label{Eq:def_s}
s_\alpha=b_{\alpha} \times \sigma_8 \, G(z) \; ,
\ee
and the effective RSD parameter is given by:
\be
\label{Eq:def_f_s}
f_s = f \times \sigma_8 \, G(z) \; .
\ee
We will employ the subscripts $R$, $E$ and $Q$
to refer to the LRG-like tracer, the ELG-like tracer, and the 
QSO-like tracer respectively.

Our fiducial model is a standard, flat $\Lambda$CDM model with $\Omega_m=0.27$,
$h=0.7$, $n_s=0.96$, $\Omega_\nu =0$ and $f_{NL}=0$. In this model, 
the RSD parameter is very well approximated by 
$f(z) = - d\log G(z)/d z  \simeq \Omega_m^{0.55}(z)$.

Our set of parameters is the following:
\be
\label{Eq:pars}
\theta^i = \{ \log H(z) , \log D_A(z) , s_R (z), s_E(z), s_Q(z), f_s(z), f_{NL}(z) \} \; .
\ee
In this paper we will discuss only the {\em conditional} errors (i.e. the reciprocals of the
Fisher matrix elements), hence we have not 
included any global cosmological parameters such as $\Omega_m$, $h$, or $n_s$
in our analysis. It only makes sense to include global parameters, and their
covariance with the parameters on each redshift slice, when we
also consider priors. However, this would obfuscate the contribution that arises
from the relative clusterings between the tracers, so we leave this important
issue to a future paper.

\begin{figure}
\includegraphics[width=80mm]{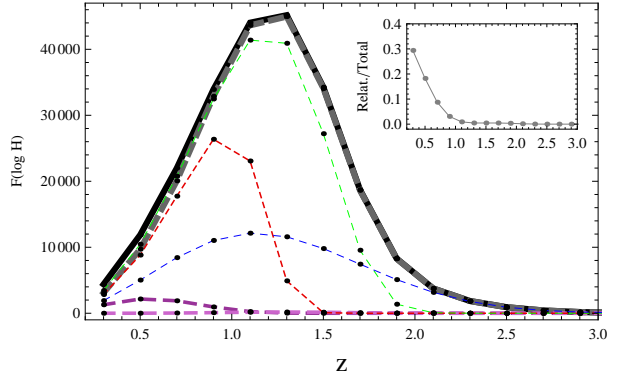}
\caption{Fisher matrix element $F(\log H,\log H)=H^2/\sigma_c^2(H)$, where
$\sigma_c(H)$ are the {\em conditional}  errors on $H$.
The thick solid (black) line corresponds to the total Fisher matrix, including 
information from all of the tracers as well as their cross-correlations.
The thin short-dashed lines
correspond to the Fisher information obtained by considering only the 
individual tracers (in the color version, red, green and blue correspond to 
LRGs, ELGs and QSOs, respectively). The thick long-dashed line 
(gray in color version) corresponds to the Fisher information 
associated with the total effective volume of Eq. \eqref{Eq:f_1}, and is subjected to the limitations imposed by 
cosmic variance. The long-dashed lines (dark and light purple in color version) 
correspond to the
the contributions to the Fisher matrix which come from the relative clusterings
between the different tracers, shown in Eq. \eqref{Eq:f_a}.
The inset shows the contribution of the relational information to the total information
on each redshift slice.
}
\label{Fig:F_lnH}
\end{figure}

\begin{figure}
\includegraphics[width=80mm]{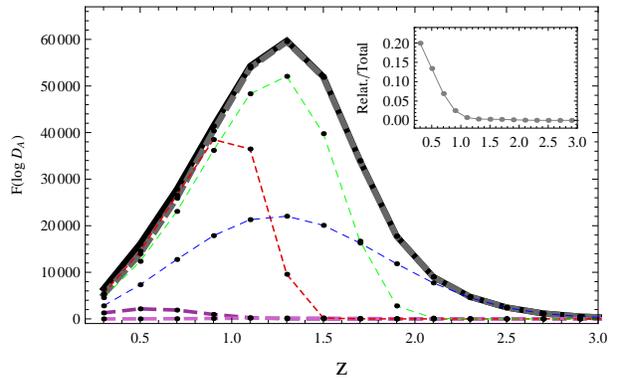}
\caption{Fisher matrix element $F(\log D_A,\log D_A)=D_A^2/\sigma_c^2(D_A)$.
The lines and the inset follow the same prescription as described in the caption for Fig. \ref{Fig:F_lnH}.
}
\label{Fig:F_lnDa}
\end{figure}

In Figs. \ref{Fig:F_lnH} and \ref{Fig:F_lnDa} we show the Fisher information for 
$\log H(z)$ and $\log D_A(z)$ on each redshift slice. 
These Fisher matrix elements
are related to the conditional errors by $F(\log H,\log H)=H^2/\sigma_c^2(H)$, and similarly for $\log D_A$ ($\sigma_c$ are the uncertainties assuming that all other parameters are 
kept fixed). The information in the first few redshift slices is smaller simply because the 
volume of the survey grows very fast with redshift.

\begin{figure}
\includegraphics[width=80mm]{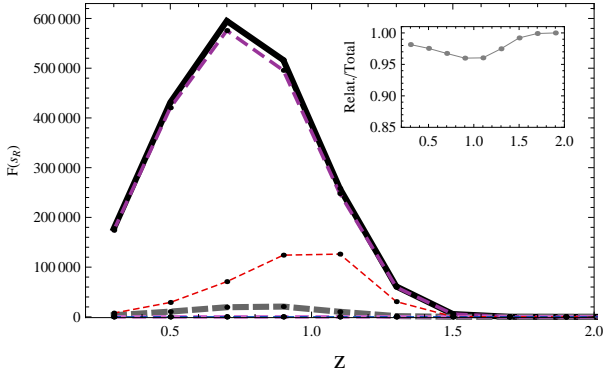}
\caption{Fisher matrix element $F(s_R,s_R)=s_R^2/\sigma_c^2(s_R)$
for the effective bias of the LRG-like tracer, 
$s_R=b_{R} \times \sigma_8 G(z)$.}
\label{Fig:F_bLRG}
\end{figure}

\begin{figure}
\includegraphics[width=80mm]{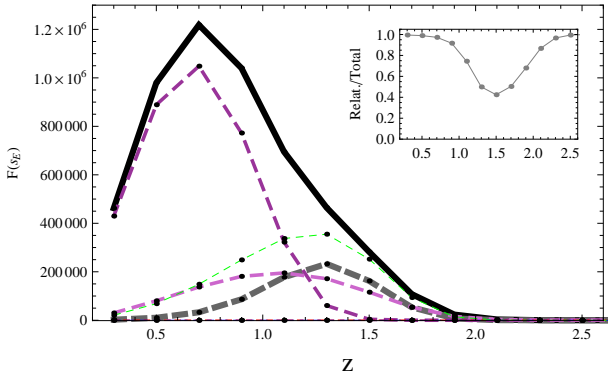}
\caption{Fisher matrix element $F(s_E,s_E)=s_E^2/\sigma_c^2(s_E)$
for the effective bias of the ELG-like tracer, 
$s_E=b_{E} \times \sigma_8 G(z)$. 
}
\label{Fig:F_bELG}
\end{figure}

\begin{figure}
\includegraphics[width=80mm]{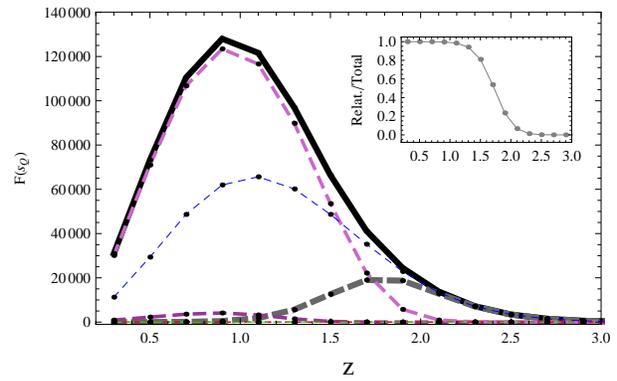}
\caption{Fisher matrix element $F(s_Q,s_Q)=s_Q^2/\sigma_c^2(s_Q)$
for the effective bias of the QSO-like tracer, 
$s_Q=b_{Q} \times \sigma_8 G(z)$.
}
\label{Fig:F_bQ}
\end{figure}

\begin{figure}
\includegraphics[width=80mm]{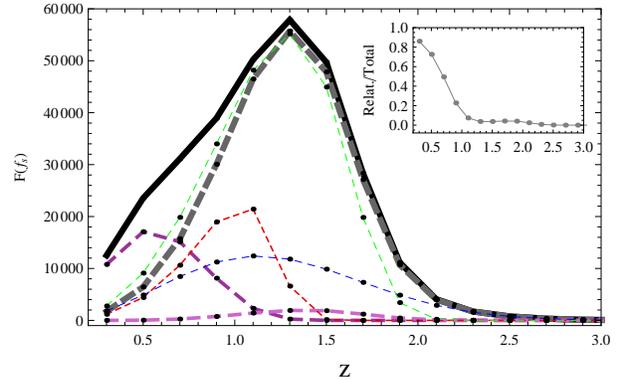}
\caption{Fisher matrix element for the effective redshift distortion parameter,
$f_s(z)= f(z) \, \sigma_8 \, G(z)$.
}
\label{Fig:F_fs}
\end{figure}

\begin{figure}
\includegraphics[width=80mm]{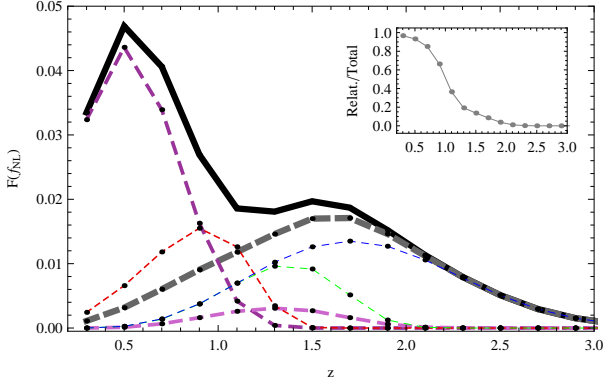}
\caption{Fisher matrix element for local non-Gaussianity parameter $f_{NL}$.
}
\label{Fig:F_fNL}
\end{figure}

In Figs. \ref{Fig:F_lnH}-\ref{Fig:F_fNL}, the thin short-dashed lines correspond
to the Fisher information obtained by using only the LRG-like tracer (red in the
color version), only the ELG-like tracer (green), or only the QSO-like tracer (blue). 
The thick solid line shows the Fisher information obtained by
using all the tracers, as well as their cross-correlations; it is always
above the Fisher information from the individual tracers. The thick long-dashed 
line (grey in color version) corresponds to the Fisher information which comes
from the total effective volume of Eq. \eqref{Eq:f_1}, and is therefore the
quantity which is subject to cosmic variance. The long-dashed lines
(dark and light purple in color version) correspond to the information in
the relative clusterings between the three species of tracers, and
are not subject to the same bounds as the effective volume -- see Eq. \eqref{Eq:f_a}. 
The sum of those three contributions (the thick long-dashed lines) is 
exactly equal to the total Fisher information
(denoted by the thick solid line).

The insets (upper right corners) in Figs. \ref{Fig:F_lnH}-\ref{Fig:F_fNL}
show the fraction of the Fisher information contributed
by the relative clusterings between the tracers. As one can see,
there is a $\sim 30\%$ enhancement in the information of $H(z)$, and
a $\sim 20\%$ enhancement in the information of $D_A(z)$, 
which comes from the information in the relative amplitudes of the clusterings. 
These enhancementes can be understood as follows: although 
the prefactor of Eq. \eqref{Eq:ObsPk} contains most of the information about 
$H(z)$ and $D_A(z)$,
that coefficient cancels out from the ratios between the effective powers, and therefore
does not contribute anything to the Fisher information of the relative clusterings.
However, the observed spectra also depend on $H(z)$ and $D_A(z)$ through 
the radial and angular components of the Fourier modes [see Eq. \eqref{Eq:ks}],
and that leads to some amount of information about BAOs in the relative clusterings. 
In fact, by increasing the redshift errors to $\sigma_z > 0.01 (1+z)$ we lose the ability
to measure anisotropies in the clustering, and as a consequence the Fisher 
information from the relative clusterings to $H(z)$ and $D_A(z)$ becomes negligible.

In the case of the biases of the tracers, there is a dramatic improvement
coming from the relative clusterings. In Figs. \ref{Fig:F_bLRG}, \ref{Fig:F_bELG}
and \ref{Fig:F_bQ} we show the Fisher information for the effective biases of the 
LRG-, ELG-, and QSO-like tracers, respectively.
For the biases, the information arising from the relative clusterings is almost
always dominant with respect to the information which comes from the effective volume.
Moreover, including other species of tracers with different biases
improves the information in the original bias by an order of magnitude, compared
with the single-tracer analysis. There is a caveat, though: including other
species of tracers also introduces cross-correlations between the 
measurements of the biases; so the conditional errors implied by the information
shown in Figs. \ref{Fig:F_bLRG}-\ref{Fig:F_bQ} are 
substantially underestimated with respect to the actual, marginalized uncertainties
in those biases.

The most significant results are those for the effective RSD parameter, 
$f_s(z)= f(z) \, \sigma_8 \, G(z)$, and for the local non-Gaussian parameter,
$f_{NL}$. We model the non-Gaussianities in the usual way \citep{NGDalal},
through a bias correction: 
\be
\label{Eq:NG}
\Delta b_{NL} = f_{NL} (b-1) \times 
\frac{3 \, \delta_c  \, \Omega_m  \, H_0^2}{c^2  \, k^2  \, T(k)  \, G(z)} \; ,
\ee
where $\delta_c=1.68$ is the critical linear density for spherical collapse, and $T(k)$
is the matter transfer function normalized to unity at large scales ($k=0$).
This may not be a good approximation for quasars \citep{Slosaretal08}, for which 
the factor $(b-1)$ in Eq. \eqref{Eq:NG} should perhaps be substituted by $(b-1.6)$, 
but we have kept the original definition for the sake of comparison between tracers 
of different biases.

In Fig. \ref{Fig:F_fs} we show the Fisher information for $f_s$ on each
redshift slice, and it is clear that the information from relative clusterings
can enhance the information in the effective volume by up to a factor of $\sim$ 10. 
The largest improvement is precisely at the lowest redshift slices, which are also 
the ones which are most sensitive to dark energy and
modified gravity models. This figure shows that, in order to measure the
RSD parameter with high accuracy (both as a function and as a function of
scale), we need to observe many more tracers at redshifts $z\lesssim 1$ 
than a simple argument based on shot noise would indicate.
These results confirm and extend those of \citet{Hamaus2012}.

The Fisher information for $f_{NL}$ is shown in Fig. \ref{Fig:F_fNL}.
There is a dramatic enhancement in the information at low redshifts,
with respect to what we expect solely from the effective volume. In fact, 
the improvement is so large that it may seem at odds with the fact that
non-Gaussianities are manifested through bias-dependent modifications to
the transfer function on large scales. However, for the scenario that we 
considered in this Section, there is so much more information {\em density} 
in the low redshift slices (where there is a high density of tracers), 
which come from comparing the clusterings between the different tracers,
that this more than compensates for the relatively small scales spanned by 
those low-$z$ slices. The bottom line is that it is possible to enhance the 
constraints on $f_{NL}$ by large factors. This is also consistent with 
the findings of \citet{Hamaus2011}.

\section{Conclusions}
\label{S:4}

We have shown how galaxy surveys can constrain
bias-sensitive parameters to an accuracy which is
not limited by cosmic variance. This emerges naturally from the
multi-tracer Fisher information matrix, whose eigenvalues include both
the effective volume (which is limited by cosmic variance) and the information 
from comparing the relative amplitudes of clustering
between tracers. It is trivial to employ the methods described in this 
paper to forecast the power of any survey, and to any number of tracers of large-scale
structure. The task of producing forecasts is further simplified due 
to the diagonalized form of the multi-tracer Fisher matrix.

In order to promote clarity we made several simplifying, but probably 
over-optimistic assumptions: e.g., we took the shot noise of auto-correlations 
to be Poissonian, but assumed that there is no shot noise for the cross-correlations
between different tracers. We also assumed that the different tracers of
large-scale structure correspond to non-overlapping halo masses, so that 
the biases are not correlated -- and, as shown by \citet{GM2010,CaiCV}, these
correlations limit the benefits of the multi-tracer approach.
In the context of the formalism presented in this paper, we can go beyond 
these simple assumptions by considering a more realistic 
covariance for the counts in cells, and then computing the corresponding multi-tracer 
Fisher information matrix, as was carried out in \citet{AbramoFisher}. 
A particularly useful extension would be to compute the optimal
binning for the bias in terms of the distributions of halo masses, 
including the cross-correlations between the biases of the different tracers.

Moreover, because priors and marginalizations tend to obfuscate the
different contributions that determine each constraint,
we have decided to focus on the Fisher information itself, which is 
related to the conditional errors by $\sigma_c(\theta^i)=1/\sqrt{F_{ii}}$. 
However, it is clear that considering several different types of tracers also
brings additional covariances, so the actual constraints (after all
the marginalizations and priors) will not be enhanced by the same factor.
A crucial next step to our analysis would be to consider the marginalized 
constraints subject to some suitable sets of priors. In particular, as
already shown by \citet{CaiWL}, the role of
lensing as a means to determine priors on the biases of different
tracers of large-scale structure seems a promising way to mitigate
some of the covariances inherent to multi-tracer surveys.

Our analysis shows that the parameters whose constraints have the most
to gain from a multi-tracer survey are the RSD parameter, $f(z)$, and the local 
non-Gaussianity parameter, $f_{NL}$. For the example employed in Section \ref{S:3}, 
the Fisher information of the effective RSD parameter increases by a factor of 
up to $\sim 10$ when the information from relative clustering is included, 
which means a factor up to $\sim 3$ reduction in the conditional errors for 
that parameter. For $f_{NL}$,
the Fisher information is boosted by an even more dramatic factor, of $\sim 20$.
For the hypothetical survey and tracers that we considered, the largest gains 
would occur at low redshifts, because of the higher number densities of objects
in the low-$z$ slices.

Our analysis also shows that it is possible to obtain enhancements 
in the determination of radial and angular BAOs from the relative clusterings: in
the scenario we considered, these enhancements were up to
$\sim 20\%$ for the Fisher information of the angular distance $D_A(z)$, and
up to $\sim 30\%$ for the Fisher information of $H(z)$. The explanation for these
enhancements comes from the fact that radial and angular BAOs are also 
manifested in the anisotropies of the clustering, through $\mu_k$ -- see Eq. 
\eqref{Eq:ks}. This is in fact a multi-tracer Alcock-Paczynski test
\citep{1979Natur.281..358A}, hence it depends on whether or not the redshift 
accuracy of the survey allows the measurement of anisotropies in the 
redshift-space clustering.
Indeed, we have checked that for redshift errors $\sigma_z \gtrsim 0.01 (1+z)$ there
is basically no additional contribution from the relative clusterings to the Fisher
informations of $D_A(z)$ and $H(z)$.

The key point in this paper is that whenever a redshift survey
detects tracers of large-scale structure whose clustering amplitudes
are different, comparisons between the clusterings tap into additional 
sources of information about the parameters that describe them. 
It may be useful to think of a ``generalized bias'' for each tracer, of the form
$B_\alpha(z;k,\mu_k) = b_\alpha (z;k) + f_2(z;k) \mu_k^2 + \ldots $ ,
which includes as many parameters and dependencies as one may wish to consider.
As long as there are sufficiently high densities of tracers with distinct (i.e., 
non-degenerate) biases, we can measure $B_\alpha (z;k,\mu_k)$ in bins of 
$z$, $k$ and $\mu_k$ to an accuracy which is, in principle, not limited by the 
volume of the survey.

\vskip 0.5cm

\noindent {\it  Acknowledgements} -- 
R.A. would like to thank FAPESP and CNPq, and 
K. L. would like to acknowledge the support of the Sociedade Brasileira de 
F\'{\i}sica (SBF) and the American Physical Society (APS).

\bibliographystyle{mn2e.bst}

\end{document}